\documentclass{osa-article}

\usepackage{upgreek}
\usepackage{gensymb}
\usepackage{tablefootnote}
\usepackage{threeparttable}
\usepackage{placeins}
\usepackage[T1]{fontenc}
\pdfoutput=1

\definecolor{darkgreen}{rgb}{0.0, 0.4, 0.0}

\journal{osajournal}

\articletype{Research Article}

\begin{document}

\title{Single-Shot Optical Neural Network}

\author{Liane Bernstein\authormark{1,3}, Alexander Sludds\authormark{1}, Christopher Panuski\authormark{1}, Sivan Trajtenberg-Mills\authormark{1}, Ryan Hamerly\authormark{1,2}, and Dirk Englund\authormark{1}}

\address{\authormark{1}Research Laboratory of Electronics, Massachusetts Institute of Technology, 50 Vassar St, Cambridge, MA~02139, USA\\
\authormark{2}NTT Research, Inc., Physics \& Informatics Laboratories, Sunnyvale, CA~94085, USA\\
\authormark{3}lbern@mit.edu}


\begin{abstract*}
As deep neural networks (DNNs) grow to solve increasingly complex problems, they are becoming limited by the latency and power consumption of existing digital processors. For improved speed and energy efficiency, specialized analog optical and electronic hardware has been proposed, however, with limited scalability (input vector length $K$ of hundreds of elements). Here, we present a scalable, single-shot-per-layer analog optical processor that uses free-space optics to reconfigurably distribute an input vector and integrated optoelectronics for static, updatable weighting and the nonlinearity --- with $K\approx 1,000$ and beyond. We experimentally test classification accuracy of the MNIST handwritten digit dataset, achieving 94.7\% (ground truth 96.3\%) without data preprocessing or retraining on the hardware. We also determine the fundamental upper bound on throughput ($\sim$0.9~exaMAC/s), set by the maximum optical bandwidth before significant increase in error. Our combination of wide spectral and spatial bandwidths in a CMOS-compatible system enables highly efficient computing for next-generation DNNs.
\end{abstract*}

\section{Introduction}

Artificial deep neural networks (DNNs) have revolutionized automated image classification~\cite{krizhevsky_imagenet_2012}, natural language processing~\cite{wolf_transformers_2020} and medical prediction and diagnosis~\cite{esteva2019guide}. These breakthroughs were made possible by sufficiently large datasets and computing capacity, since an increase in DNN model size tends to yield higher accuracy~\cite{Xu2018-pj,kaplan2020scaling}. Recently, though specialized digital hardware has significantly improved the latency, energy consumption and throughput of deep learning~\cite{nurvitadhi2017can,jouppi_tpu}, the computation and memory costs of modern DNNs have continued to grow exponentially, outpacing the Moore's law increases in microprocessor performance~\cite{Xu2018-pj}. Inference, which accounts for 80-90\% of machine learning tasks in data centers~\cite{patterson2021carbon}, has particularly stringent requirements for low latency, such as in translation and autonomous driving, as well as in newer applications in astrophysics~\cite{huerta2019enabling} and the control of fusion reactors~\cite{degrave2022magnetic}.

The rate-limiting step in DNN inference is matrix-vector multiplication (MVM), which implements the synaptic connections in a network layer: a matrix of learned weights multiplies an input vector of length $K$ (which often exceeds $K=1$,000 in modern DNN workloads~\cite{kaplan2020scaling}). Within MVM, the most expensive task is data movement of the weights and inputs from memory to the multiplier units, followed by the multiply-accumulate (MAC) operation itself~\cite{jouppi_tpu,Sze2017-qg}. Static, `weight-stationary' architectures~\cite{Sze2017-qg} (e.g., in-memory computing~\cite{sebastian2020memory}) can yield significant improvements in efficiency, since in this dataflow, the input vector is broadcast across the multiplier units that apply fixed weights, and the output vector can, in theory, be read out in one step. DNN layer outputs can thus be computed in a single shot without costly weight updates \textit{if} the entire weight matrix can be stored on the hardware.

A survey of previous weight-stationary DNN accelerators~\cite{jouppi_tpu,Yao2020-cx,Xiao2020-ya,Shen2017-vq,Tait2017-in, Feldmann2021-parallel} reveals a tradeoff between latency, energy consumption, and scalability, i.e., the number of weights that the hardware can accommodate. In digital electronic circuits (e.g., Google's Tensor Processing Unit, or TPU~\cite{jouppi_tpu}), constraints on wiring for digital data transmission require partial products to be passed from one multiplier to the next at each clock cycle, which prohibits single-shot MVM. Analog electronic circuits such as memristor crossbar arrays~\cite{Yao2020-cx}, on the other hand, can sum partial products along a wire by Kirchhoff's current law, but they are bounded in speed, energy consumption and accuracy by parasitic capacitance, resistance, and current leakage~\cite{xiao2020analog}. Integrated 2D optical schemes can compute at still higher speeds~\cite{Shen2017-vq,Tait2017-in,Feldmann2021-parallel}, but limits on control, multiplexing, component area and on-chip losses through $\mathcal{O}(K)$-depth circuits constrain scalability to a vector length $K$ of hundreds of elements.

\begin{figure}[htpb]
    \centering
    \includegraphics[width=1\linewidth]{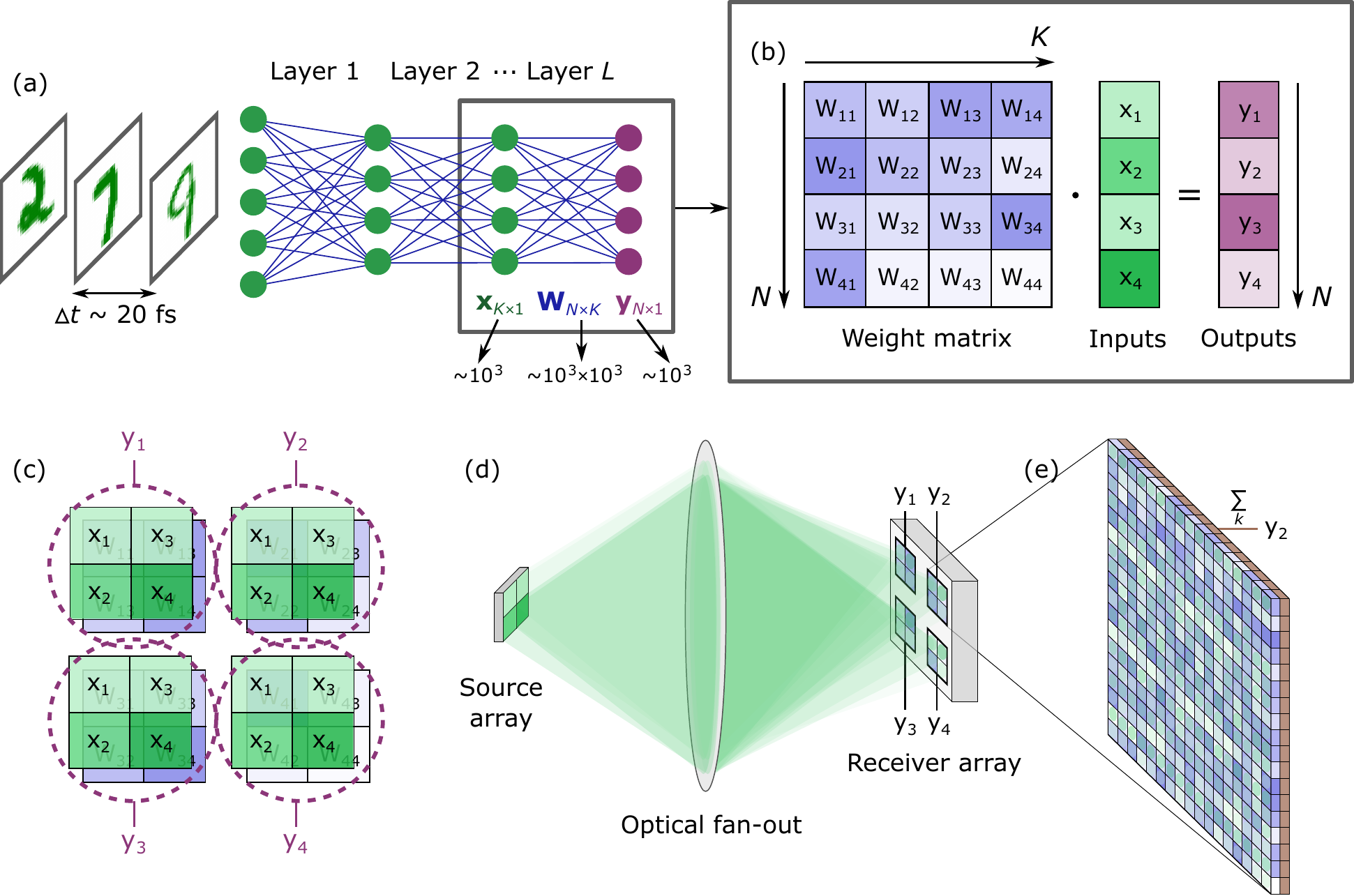}
    \caption{Single-shot computation of a fully connected neural network (FC-NN) layer. \textbf{a},~An FC-NN determines the output classes \textbf{y} of input images. \textbf{b},~One layer: weight matrix \textbf{W}$_{N\times K}$ multiplies input activation vector \textbf{x}$_{K\times 1}$ followed by a nonlinearity (e.g., Rectified Linear Unit, not shown) to produce output activation vector \textbf{y}$_{N\times 1}$. \textbf{c},~Block encoding and fan-out of \textbf{x} over rows of \textbf{W} and block-wise summation computes \textbf{W}$\cdot$\textbf{x} in one time step. \textbf{d},~$K$-element source array encodes \textbf{x} into analog intensities of light pulses, which are replicated and imaged onto $N$ receiver blocks, with electronics for summation and nonlinearity. Outputs broadcast to the next layer, e.g., a duplicate of the same hardware. \textbf{e},~Free-space optics enable high-density, 3D information transfer, with $\sim$10$^3$ inputs incident on $\sim$10$^3$ weighting elements per block above electrically connected photodetectors.}
    \label{fig:concept}
\end{figure}

\FloatBarrier

A 3D optical architecture permits greater connectivity between potentially millions of spatial modes~\cite{Miller2017-qd,maktoobi2019diffractive}, given the prevalence of megapixel-scale optoelectronic components (e.g., cameras and spatial light modulators). 3D optical MVM accelerators were among the first to be investigated by pioneers in the field~\cite{Goodman1978-yn,athale1982optical,psaltis1985optical,wagner1987multilayer} --- partly out of necessity prior to the advent of compact low-loss photonic integrated circuits --- but they did not achieve widespread adoption for two reasons. First, competition from digital electronic computers limited their application space~\cite{goodman19914}. Second, they suffered from a lack of flexibility, as experimental demonstrations employed fixed weighting masks or bulky modulators and were small in scale (input vectors with $\sim$10s of elements). Recently, theoretical analyses showing the potential of ultra-low-energy computing for DNNs (e.g.,~\cite{hamerly2019-large-scale}) have driven a resurgent interest in 3D optical processors; however, experimental realizations have been limited to vector-vector dot products~\cite{wang2022-1photon}, MVM at small scale ($K = 56$ elements)~\cite{spall2020fully}, and diffractive~/ convolutional models with limited matrix expressivity~\cite{Zhou2021-diffractive,miscuglio2020massively,Lin2018-qh,chang2018hybrid}.

Here, we present for the first time a demonstration of a fully programmable, 3D optical neural network (ONN) capable of single-shot-per-layer inference at large scale. Optical sources encode an input vector and free-space optical components copy and distribute this display (\textit{fan-out}) to updatable optoelectronic weighting elements. Electronics sum the partial products and implement the nonlinearity. We thus combine the advantages of free-space optics (tightly packed spatial modes) and integrated optoelectronics (reconfigurable weighting and a low-power nonlinearity) to enable scalability. We show high classification accuracy of the MNIST handwritten digit dataset in a proof-of-concept experiment with $K=784$, where dynamic reconfigurability of the inputs, fan-out and weighting elements allow for potential model updates. We then determine the physical limit to throughput of our system ($\sim$0.9~exaMAC/s) by measuring classification accuracy versus source spectral width. Lastly, we calculate the theoretical total latency ($\sim$10~ns), energy consumption ($\sim$10~fJ/MAC) and throughput ($\sim$petaMAC/s) to process a complete million-element DNN layer (with $K=1,$000) with near-term, CMOS-compatible technologies. This highly efficient architecture can enable next-generation DNNs while also having a significant impact on other fields such as non-convex optimization (e.g., Ising problems), signal processing and other machine-learning tasks where MVM also dominates energy consumption and latency.

\begin{figure}[htpb]
    \centering
    \includegraphics[width=.8\linewidth]{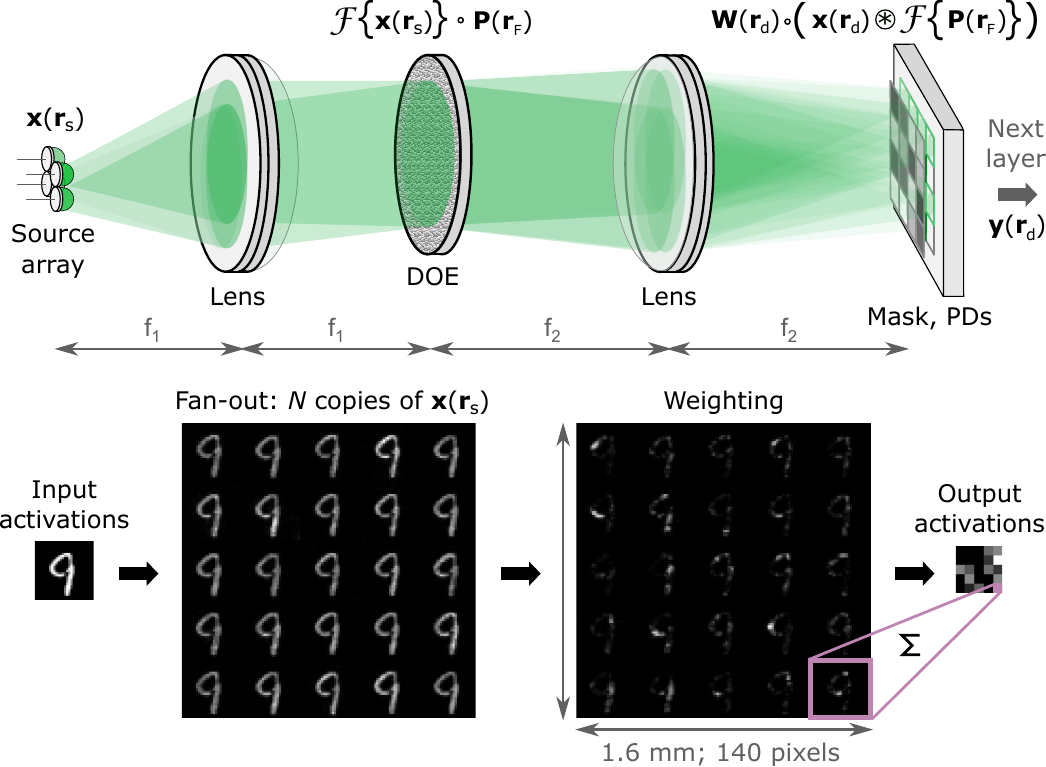}
    \caption{Single-shot optical neural network. Source array (wavelength $\lambda$, object plane) encodes inputs $\textbf{x}(\textbf{r}_\text{s}$) into analog optical intensities at transverse spatial positions $\textbf{r}_\text{s}$. Diffractive optical element (DOE, Fourier plane) performs element-wise multiplication of the spatial Fourier transform of $\textbf{x}(\textbf{r}_\text{s})$ with fan-out phase pattern $\textbf{P}(\textbf{r}_\text{F})$, where $\textbf{r}_\text{F} = 2\uppi\cdot \textbf{r}_\text{s}/(\lambda\cdot\text{f}_1$) and `F' subscript denotes Fourier plane. Optoelectronic weighting elements (image plane) perform element-wise products between weight matrix and replicated input activations: $\textbf{W}(\textbf{r}_\text{d})\circ\left(\textbf{x}\left(\textbf{r}_\text{d}\right)\circledast\mathscr{F}\left\{ \text{\textbf{P}}\left(\textbf{r}_\text{F}\right)\right\}\right)$, where $\textbf{r}_\text{d}=-(\text{f}_1/\text{f}_2)~\textbf{r}_\text{s}$ and `d' subscript denotes detector plane. Electronics sum $K$ photodetector outputs per block by Kirchhoff’s current law. Experimental fan-out and weighting data shown.}
    \label{fig:schematic}
\end{figure}


\section{Concept}

As illustrated in Fig.~\ref{fig:concept}, the key feature of our architecture is the ability to compute a DNN inference layer in a single shot. In a fully connected neural network (Fig.~\ref{fig:concept}a,b), a layer's output elements (\textbf{y}) are the weighted sums of the inputs (\textbf{x}), cascaded into a nonlinear activation function ($f$) that approximates the thresholding operation in the brain~\cite{Sze2017-qg}. In other words, a fully connected layer is equivalent to an MVM, with $\text{\textbf{y}}_{N\times 1} = f \left( \text{\textbf{W}}_{N\times K}\cdot \text{\textbf{x}}_{K\times 1}\right)$, where the weights (\textbf{W}) have been pre-learned during the training phase. Convolutional layers can also be implemented by matrix multiplication, e.g., using a Toeplitz matrix~\cite{Sze2017-qg}. In Fig.~\ref{fig:concept}c, \textbf{x} and the rows of \textbf{W} are both recast into blocks, with \textbf{x} projected onto each row of \textbf{W}. In the case of a 2D input image, \textbf{x} is already in block form. With the inputs then local to their corresponding weights, all required element-wise multiplications are completed simultaneously. Free-space optics can passively realize this data routing, replication and weighting (Fig.~\ref{fig:concept}d,e).

In our proposed architecture, a 4$f$ lens system performs the single-shot MVM (Fig.~\ref{fig:schematic}). A source array encodes the input activations into the single-spatial-mode analog amplitudes of light pulses in the object plane $\textbf{x}(\textbf{r}_\text{s}$), where \textbf{r$_\text{s}$} is the transverse spatial position in the 2D source plane. These sources can operate at GHz rates using vertical-cavity surface-emitting lasers (VCSELs)~\cite{skalli2022photonic}, $\upmu$LEDs~\cite{wong2019progress} or emerging high-speed spatial light modulators (SLMs)~\cite{panuski2022slm}. In the Fourier plane, a diffractive optical element (DOE), e.g., a liquid crystal on silicon (LCoS) SLM, displays a spot array generation phase pattern for reconfigurable fan-out. Following Fourier convolution theory, the spot array in the image plane generated by the DOE is optically convolved with the image of the input pattern, which yields $N$ copies of the input pattern. Reconfigurable weighting elements, e.g., LCoS SLM pixels plus a polarizer, then attenuate the intensity of each replicated input pixel proportionally to each weight value in \textbf{W}. $\upmu$m-scale photodetectors (PDs)~\cite{fahs2017design, zimmermann2009blue} directly below the weighting elements can convert the signal to analog electronics for block-wise summation by Kirchhoff's current law. With `optical fan-in'~\cite{wang2022-1photon}, a single large PD can replace the block of small, electrically connected PDs. An amplifier per block reads out the accumulated charge and an electronic post-processing unit performs the nonlinearity to yield \textbf{y}. An output source array (e.g., same components as the input sources) with one source per block can then be the input to the next layer.

\section{Results}

\begin{figure}[htp]
    \centering
    \includegraphics[width=1\linewidth]{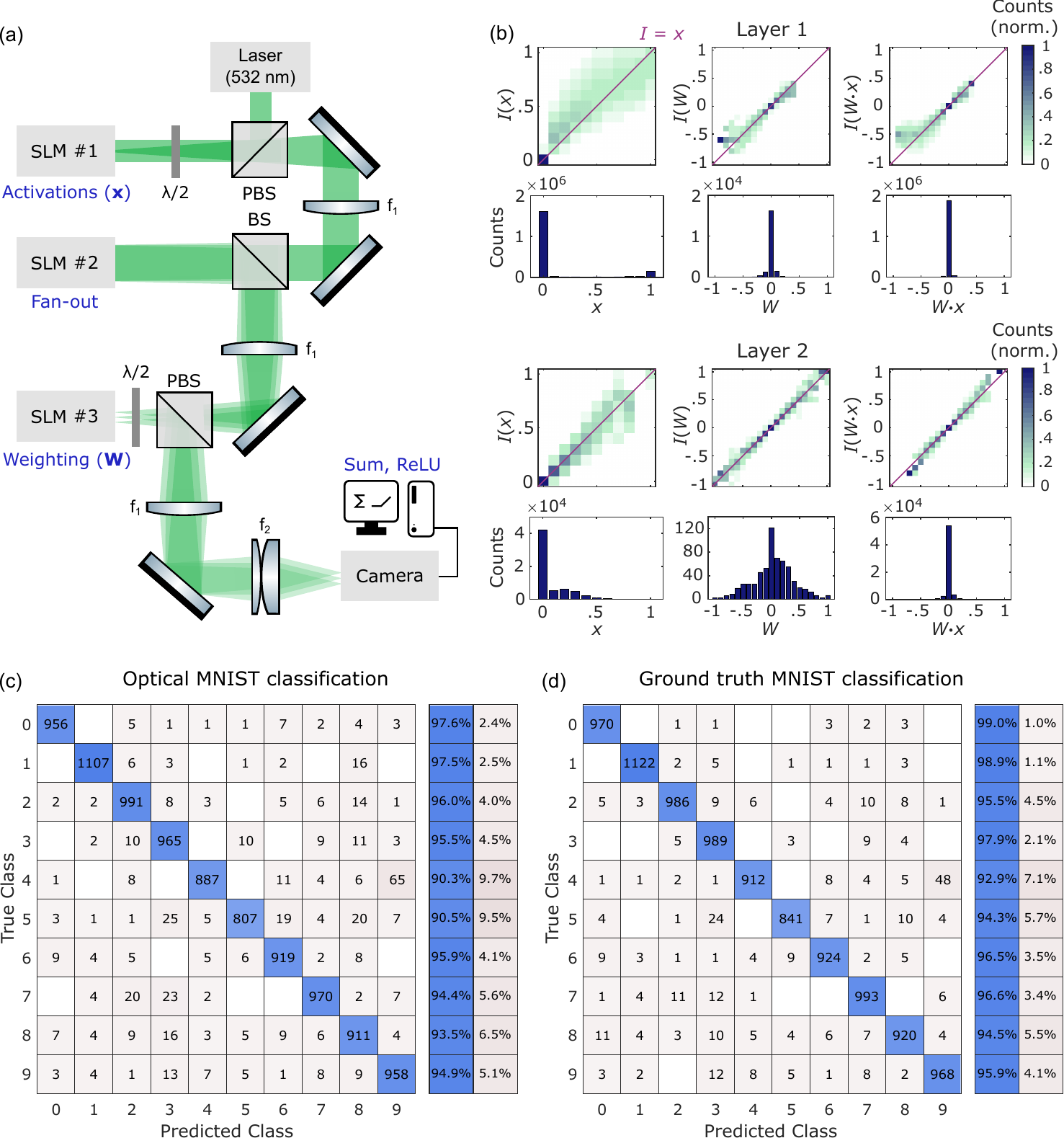}
    \caption{Proof-of-concept implementation of single-shot optical neural network. \textbf{a},~Collimated laser light (532~nm) incident on spatial light modulator (SLM~\#1, object plane) with polarization of 45\degree~after half-wave plate ($\uplambda$/2). SLM~\#1 with polarizing beamsplitter (PBS) encodes \textbf{x} by pixel-wise intensity modulation. SLM~\#2 (Fourier plane) imparts fan-out phase pattern. Achromatic lenses of focal lengths $f_1=250$~mm and $f_2=145$~mm image replicated \textbf{x} from SLM~\#1 to SLM~\#3 for weighting (\textbf{W}) and from SLM~\#3 to camera. A digital computer controls hardware, sums each block and implements nonlinearity. \textbf{b},~Histograms of received intensities $I$ (normalized) versus transmitted values over all pixels for activations $x$, weights $W$ and element-wise products $W\cdot x$ from 100 randomly selected MNIST test images for first and second layers of the DNN. Each column normalized by sum of column (`Counts' in 1D histograms). \textbf{c-d},~Confusion matrices for classification of 10,000 previously unseen MNIST test images by our optical system (\textbf{c}, 94.7\% accuracy) and by a standard digital computer (\textbf{d}, 96.3\% accuracy).}
    \label{fig:characterization}
\end{figure}

We experimentally verified the impact of single-shot analog optical data encoding, fan-out and weighting on the classification accuracy of the MNIST handwritten digit dataset~\cite{lecun1998gradient}. We used a two-hidden-layer fully connected neural network (FC-NN) with $784 \rightarrow 25 \rightarrow 25 \rightarrow 10$ activations (see Methods for the training procedure on a digital electronic computer). Figure~\ref{fig:characterization}a provides a schematic representation of our setup, where three LCoS SLMs display a complete activation image (SLM~\#1), fan-out pattern (SLM~\#2) and full weight matrix (SLM~\#3) for $28\times 28\times 25 = 19,600$ multiplications per frame (see Methods and Supplementary Note~1 for details on the optical system). As a stand-in for an analog electronic circuit, a digital electronic computer performs the per-block summation and nonlinearity.

In Fig.~\ref{fig:characterization}b, we show histograms of measured intensities $I$ versus transmitted inputs \textit{x}, weights \textit{W} and element-wise products $W\cdot x$. To evaluate $I(x)$, we displayed a different MNIST test image at each time step for 100 steps on SLM~\#1, set all SLM~\#3 pixels to a constant, uniform value, and recorded the outputs. Similarly, to measure $I(W)$, we displayed the layer weights on SLM~\#3 and set all SLM~\#1 input pixels to a constant, uniform value. For $I(W\cdot x)$, SLM~\#1 encoded 100 MNIST test images while SLM~\#3 displayed the weight matrix. The distribution $I(W\cdot x)$ versus $W\cdot x$ for layer 1 is broad for $W\cdot x \in [-1, -0.5]$, but these columns only represent 0.06\% of the total counts. Otherwise, $I(W\cdot x)$ is close to the diagonal, i.e., $I(W\cdot x)=W\cdot x$. The distributions for layer 2 are narrower than those for layer 1 due to increased pixel spacing made possible by fewer input activations (see Supplementary Note~1), which lowers crosstalk. 

Figure~\ref{fig:characterization}c,d shows confusion matrices for inference on 10,000 previously unseen MNIST test images with the two-hidden-layer FC-NN described above. We computed each layer consecutively on the hardware, where camera outputs were fed back to SLM~\#1 as inputs to the next layer. Thus, the fan-out pattern on SLM~\#2 was held constant for the duration of the experiment, the weights on SLM~\#3 were updated only between layers, and the inputs on SLM~\#1 were refreshed at every time step with each new input vector. We obtained a 94.7\% classification accuracy (Fig~\ref{fig:characterization}c), compared with the ground-truth all-electronic accuracy of 96.3\% (Fig.~\ref{fig:characterization}d). Our optical hardware's classification error was $\sim$1\% higher than the ground truth for most digits, but up to $\sim$4\% higher for the digit `5' (additionally misclassified `5's were mostly labeled `6' or `8' by our system, possibly due to blurring from crosstalk). The top-2 optical accuracy was 98.1\%, similar to the ground-truth top-2 accuracy of 98.7\%.

\begin{figure}[htpb]
    \centering
    \includegraphics[width=1\linewidth]{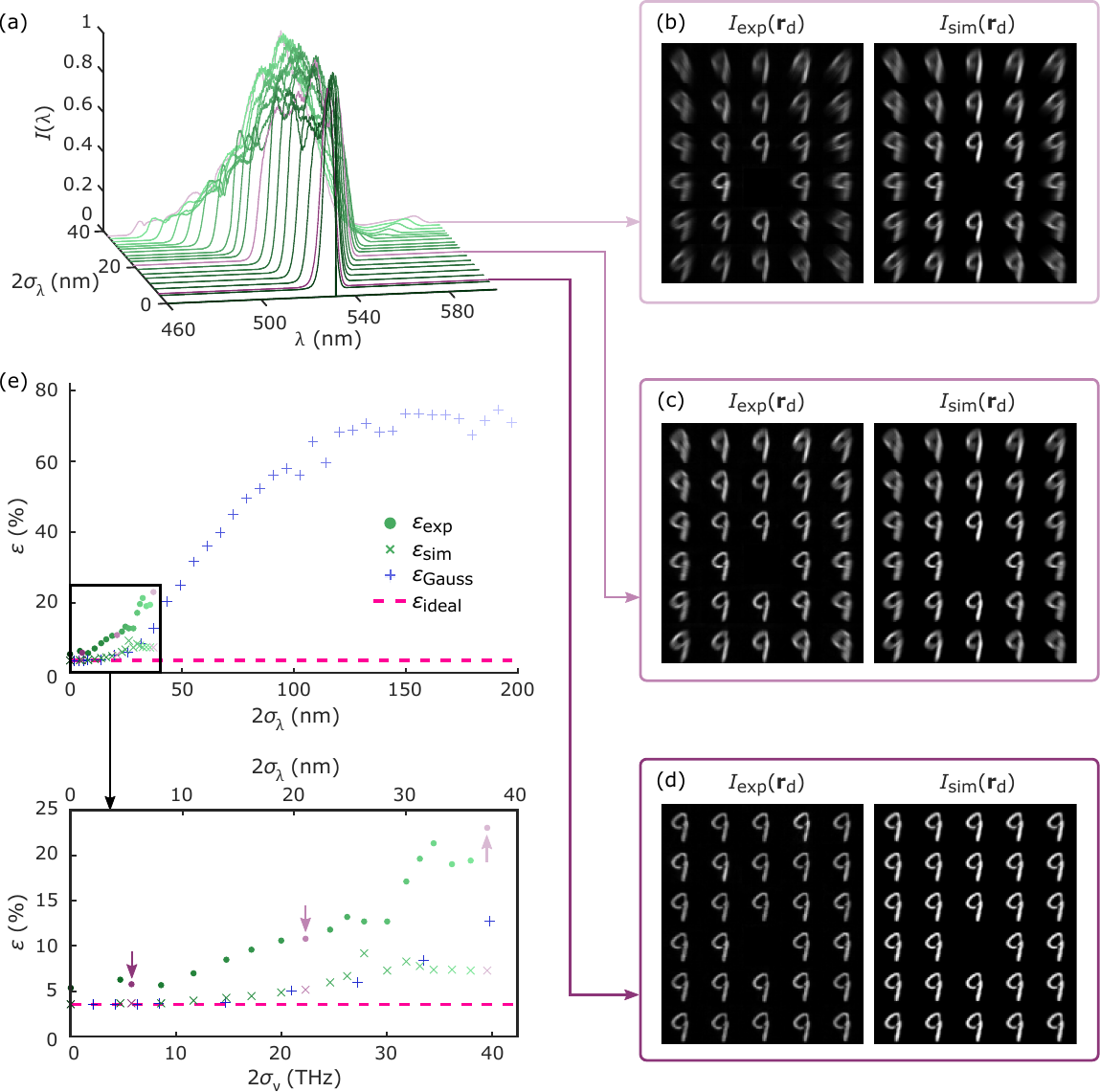}
    \caption{Maximum optical throughput determined by MNIST classification with broad-bandwidth laser as input to setup shown in Fig.~\ref{fig:characterization}a. \textbf{a},~Filtered supercontinuum laser spectra used in experiment. \textbf{b}-\textbf{d},~Example blurred images from spectral widths of 37~nm, 21~nm and 5.4~nm with measured classification errors of 23\%, 11\% and 5.8\%, respectively. Left images measured on camera; right images simulated from measured spectra. Element [4,3] overlaps with zero order from SLM~\#2 and is replaced with element [6,4] in DNN experiments. \textbf{e},~Classification error of 1,000 previously unseen MNIST test images versus spectral width (2$\times$ RMS width, i.e., 2$\sigma_\lambda$) measured experimentally ($\epsilon_\text{exp}$) and simulated from source ($\epsilon_\text{sim}$) and Gaussian spectra ($\epsilon_\text{Gauss}$); ground truth error without blurring for reference ($\epsilon_\text{ideal}$). Arrows indicate measurements from spectra shown in purple (2$\sigma_\lambda$~=~37~nm, 21~nm and 5.4~nm).}
    \label{fig:throughput_experiment}
\end{figure}

Next, we investigated the optical limit to throughput of our system. The clock rate is fundamentally limited by source bandwidth, where a broader laser can produce shorter pulses. But a wide spectrum yields blurred outputs: by the spatial Fourier transform relationship between the Fourier and image planes, the distance between each replicated input pattern and the optical axis is linear in wavelength (see Methods). Therefore, the classification error increases with bandwidth, which we verified experimentally by repeating the inference experiment described above with a tunable supercontinuum source in the place of the CW laser (Fig.~\ref{fig:throughput_experiment}, measured and simulated errors $\epsilon_\text{exp}$ and $\epsilon_\text{sim}$, respectively). We classified 1,000 images from the MNIST test set with the same two-hidden-layer DNN at different spectral widths (2$\sigma_\lambda$: twice the RMS width of the spectrum, i.e., twice the standard deviation, to account for the spectrum's irregular shape). Since our system supports a limited wavelength band, we also simulated the error that would result from Gaussian spectra with broader bandwidths ($\epsilon_\text{Gauss}$). Due to optical crosstalk, for a given spectrum, $\epsilon_\text{exp} > \epsilon_\text{sim}$, but $\epsilon_\text{exp}$ follows a similar trend to $\epsilon_\text{Gauss}$ (shifted to lower 2$\sigma_\lambda$). Furthermore, $\epsilon_\text{exp}$ doubles from 5.4\% with the CW diode to 11\% at 2$\sigma_\lambda = 21$~nm, which we define as the widest acceptable source bandwidth for preserved accuracy. The Fourier transform of the corresponding source spectrum yields a pulse of full width at half maximum $\sim$0.02~ps. Therefore, given a transform-limited source in an optimized implementation, the maximum throughput in the first layer is the number of multiply-accumulate operations (19,600) divided by the minimum pulse length, which yields $\sim$0.9~exaMAC/s. Emerging high-speed modulators, e.g., plasmonic electro-optic modulators~\cite{burla2019500}, slow-light silicon modulators~\cite{han2022ultra} or thin-film lithium niobate~\cite{kharel2021breaking}, could potentially allow us to achieve this near-exascale throughput.


\section{Discussion}

We have introduced a scalable optical neural network that can compute DNN layer outputs in a single shot. In our proof-of-concept experiment, we demonstrated low loss of classification accuracy on the MNIST handwritten digit dataset with analog optical data encoding, fan-out and weighting (94.7\% optical versus 96.3\% ground truth). We also used a filtered supercontinuum source to find that a minimum pulse duration of $\sim$0.02~ps maintains the classification error within a factor of two --- yielding a near-exascale throughput limit. To improve accuracy, we can reduce stray light on the detector: for example, higher fill-factor SLMs can lessen reflection from their backplanes, aberration-corrected lenses can focus spots more tightly in the image planes to lower crosstalk, and better optical coatings can decrease unwanted interference fringes. We can also use wider and deeper DNNs (i.e., more neurons per hidden layer, a greater number of hidden layers and added biases). Additionally, we plan to benchmark our system against other accelerators with standard DNNs such as LeNet~\cite{lecun1989backpropagation} and AlexNet~\cite{krizhevsky_imagenet_2012}. We obtained all above results without retraining on the hardware, which is preferred since retraining every processor on a large scale, e.g., in data centers, would be resource-intensive. However, to preserve accuracy for broader source spectra, we could modify our DNN model with simulated blurred images as inputs during the training phase.

Having shown that the single-shot ONN can have a small impact on DNN accuracy, the next step will be to demonstrate an implementation optimized for high throughput and low energy consumption. In Methods and Supplementary Note~2, we estimate the latency, throughput, energy and area of a near-term, CMOS-compatible system. Its latency is limited by the optical time of flight ($\sim$10~ns) and its full-utilization throughput is 1~million MACs per electronic clock cycle ($\sim$petaMAC/s). Table~\ref{tab:latency} contrasts these values to digital electronics and output-stationary dataflows, in which the inputs have to be streamed in over time, resulting in a latency that is two orders of magnitude greater for $K=1,$000. In other DNN accelerators that have limited scalability, matrices that do not fit onto the hardware must be fetched from memory over multiple time steps, subjecting them to the same `memory wall' that currently bottlenecks digital electronics.


In energy consumption, at $\sim$10~fJ/MAC, the optimized single-shot ONN can outperform digital electronics by 1-2 orders of magnitude, including sources, detectors, digital-to-analog converters (DAC), analog-to-digital converters (ADC), SLMs, transimpedance amplifiers (TIAs) and electronic nonlinearities (see Methods). To further reduce energy consumption, as the DOE and weights are only updated when the model changes, the fan-out and weighting LCoS SLMs could be replaced with elements that have zero static power consumption, e.g., with an array of optical phase change material cells~\cite{Dong2019widebandgap,jung2022integrated,zhang2021electrically}, a fixed phase mask or MEMS modulators~\cite{quack2019mems}. Alternatively, the LCoS SLMs can be optimized for low power consumption~\cite{panuski2022slm}. With an increase in source and detector efficiencies~\cite{Miller2017-qd}, the overall energy consumption could then reach the single-femtojoule per MAC regime. DAC and ADC costs may also be eliminated with analog electronic nonlinearities~\cite{xiao2020analog}, as analog output activations from one layer can be directly used as inputs to the next layer. The inputs to the first layer may also be in the analog domain, e.g., as they are read out from a sensor. In terms of footprint, with bulk optics, the single-shot ONN does consume more overall volume than digital electronic accelerators, but this volume does not contribute to fabrication cost and can be used for air cooling between racks. (The system can also be more compact with shorter-focal-length, aberration-corrected lenses.) The chip area ($\sim$30~mm$^2$), on the other hand, is similar to digital electronics (25-64~mm$^2$).

\begin{table}[htpb]
\centering
\footnotesize
\begin{threeparttable}
\caption{Latency and throughput scaling of different architectures for computation of one DNN layer}
\begin{tabular}{ccc} 
\hline
Dataflow & Latency\tnote{4}~~(ns) & Throughput\tnote{4,5} \\
\hline
Weight-stationary (near-term)\tnote{1} & $\sim$10 & $N\times K/1~\text{ns}\rightarrow$ $\sim$petaMAC/s \\
Weight-stationary (physical limit)\tnote{2} & $\sim$10 & 19,600/.02~ps $\rightarrow$ $\sim$exaMAC/s \\
Systolic array (e.g., TPU~\cite{jouppi_tpu})\tnote{3} & $N+K\rightarrow 2$,000 & $N\times K/1~\text{ns}\rightarrow$ $\sim$petaMAC/s \\
Output-stationary (e.g.,~\cite{hamerly2019-large-scale}) & $K\rightarrow 1$,000 & $N\times K/1~\text{ns}\rightarrow$ $\sim$petaMAC/s \\
\hline
\end{tabular}
\begin{tablenotes}
\item [1]Our proposed near-term optimized single-shot ONN (CMOS-compatible).
\item [2]From maximum throughput measurements in Results.
\item [3]In practice, $N=K=256$ (throughput $\sim$100~teraMAC/s) in the TPU due to wiring and utilization constraints~\cite{jouppi_tpu}.
\item [4]Assuming $N=K=1$,000.
\item [5]Assuming 100\% utilization.
\end{tablenotes}
\label{tab:latency}
\end{threeparttable}
\end{table}

In summary, we have presented a single-shot-per-layer inference machine and demonstrated low loss of classification accuracy with analog, reconfigurable optical matrix-vector multiplication. Coupled to its CMOS manufacturability, our ONN's extreme parallelism and low resource consumption (enabled by optics)  make it a viable, near-term candidate to overcome the compute bottlenecks of state-of-the-art electronics, at scale. Extensions may include spectral multiplexing to further increase scaling or to compute multiple layer or channel outputs simultaneously. With such an efficient DNN processor available to computer scientists, new, larger DNN models may be developed, facilitating the next generation of artificial intelligence.

\FloatBarrier

\section*{Methods}

\subsection*{Deep neural network training}
The FC-NN was trained on a standard digital computer in Python with the Keras library on a subset of the MNIST training set (50,000 images, where 10,000 images were reserved as a validation set to fine-tune the network hyperparameters and optical setup). L1 and L2 regularization with L1~=~L2~=~0.0001, 3\% Gaussian noise and 10\% dropout were applied to each layer. The nonlinear activation function on the final layer was softmax. The Adam optimizer minimized the categorical cross-entropy loss function for 30 epochs.

\subsection*{Experimental setup}
An LCoS SLM followed by a polarizing beamsplitter (PBS) displays an input image \textbf{x} by pixel-wise amplitude modulation of a collimated 532~nm continuous-wave (CW) laser diode. In the Fourier plane, a second LCoS phase mask --- calculated using the fixed-phase weighted Gerchberg-Saxton algorithm~\cite{Kim2019-gs} --- fans out the input. In the image plane, a third LCoS SLM (also followed by a PBS) displays a complete weight matrix for pixel-wise attenuation (multiplication) of the replicated inputs. The third SLM is imaged onto a camera using additional relay optics (Fig.~\ref{fig:characterization}a). The camera is connected to a digital electronic computer that controls the hardware, sums each block of weighted inputs and performs the nonlinearity (Rectified Linear Unit, ReLU). The electronic computer also takes the output pixel values that should be negatively weighted and multiplies them by $-1$, as the SLM can only apply the absolute values of the weights. In practice, this negative weighting can be implemented with an analog switch or two PDs per receiver pixel (see Supplementary Note~3). The layer outputs, which are always positive because of the ReLU nonlinearity, are then fed back to the first SLM, and the weights are updated for the next layer of computation. See Supplementary Note~4 for our image processing and one-time system calibration procedures. See Supplementary Note~1 for additional experimental details.

\subsection*{Theoretical classification accuracy with broad source}
An LCoS SLM imparts a phase shift of $\Delta\phi$ to an incident optical field, with:
\begin{equation}
    \Delta \phi (V) = \tfrac{2 \pi}{\lambda} (n_e(V)-n_o) a
\end{equation}
where $V$ is voltage, $\lambda$ is wavelength, $n_e$ is induced extraordinary refractive index (modulated by $V$), $n_o$ is ordinary refractive index and $a$ is liquid crystal thickness~\cite{rosales2017shape}. $\Delta\phi$ applies to the incident field's polarization component that is parallel to the extraordinary axis. 

Because SLMs~\#1 and \#3, in `amplitude mode', are imaged to the camera, a change in wavelength will simply affect their contrast, and the SLM calibration performed at $\lambda_0$~=~532~nm for output optical intensity versus input value will no longer be perfectly linear. For SLM~\#2, in phase mode, there is a slight loss of phase contrast for wavelengths different from 532~nm. The main source of error, however, arises because the fan-out spot pattern on the camera, \textbf{I}($\textbf{r}_\text{d})$, is the spatial Fourier transform of the pattern on SLM~\#2, $\textbf{P}(\textbf{r}_\text{F})$, with an argument of $\textbf{r}_\text{d} = 2\uppi\cdot \textbf{r}_\text{F}/(\lambda\cdot\text{f}_2)$. Therefore, at different wavelengths, though the spatial Fourier transform function remains the same, the argument changes:
\begin{equation}
    \textbf{r}_\text{d}(\lambda) = (\lambda_0/\lambda)\cdot \textbf{r}_\text{d}(\lambda_0).
    \label{eq:h}
\end{equation}
\noindent The distance between the zero order and each diffracted spot (i.e., the center of each input replica) thus increases linearly with wavelength under the paraxial approximation. The height and width of each individual input replica, on the other hand, stays constant since SLM~\#1 is imaged to the camera, and any chromatic aberration would cause slight blurring from a shift in the position of the focal plane rather than a change in magnification.

To simulate the blurring of each replicated input image due to the spatial Fourier transform relationship described above, we first measured the spectra of our supercontinuum laser at different filter bandwidth settings with a custom spectrometer available in our laboratory and corrected for our system's wavelength response (see Supplementary Note~5). For each sampled wavelength $\lambda$ in a corrected spectrum (every 0.7~nm), we calculated the expected locations of the input replicas from a linear change in magnification (of $\lambda/\lambda_0$) of the original spot pattern \textbf{I}($\textbf{r}_\text{d}(\lambda_0))$. We then set the intensity of all input replicas for the wavelength $\lambda$ to the intensity of the optical spectrum at $\lambda$. The sum of the images of predicted replicated inputs over the entire spectrum then produces the simulated blurred images, which we used as inputs to our two-hidden-layer DNN in inference on a digital electronic computer. The inputs to every layer were blurred following the same procedure. The resulting classification errors are shown in Fig.~\ref{fig:throughput_experiment}e. We also calculated the blurring and accuracy degradation from Gaussian spectra to model broader bandwidths (which our experiment did not support due to the dielectric mirror of SLM~\#2).

\subsection*{Latency, throughput, energy, area of optimized system versus other architectures}
The optimized single-shot ONN's latency of $\sim$10~ns is the summed update time of the electronic elements and the optical time of flight. In such a weight-stationary system, latency is independent of vector length as long as the weight matrix fits onto the hardware. In weight-stationary digital electronics (e.g., systolic arrays like Google's TPU~\cite{jouppi_tpu}), on the other hand, where inputs are message-passed across the weight matrix due to wiring constraints, the latency is at least $N+K$ clock cycles for an ($N\times K$)-sized MVM. Similarly, in output-stationary architectures~\cite{hamerly2019-large-scale,bernstein2021-freely}, latency scales with $K$, as the inputs are streamed in over time. If $K = N = 1\text{,}000$, our proposed near-term optical processor then outperforms these architectures by two orders of magnitude. In throughput, because operations are pipelined to compute $10^{6}$ MACs every $\sim$1~ns with 100\% utilization, the system computes $\sim$10$^{15}$~MAC/s --- a throughput on the order of petaMAC/s.

In energy consumption, digital DNN accelerators are limited primarily by data movement to $\sim$0.1-1~pJ/MAC~\cite{jouppi_tpu,Shao2019-le}, depending on the implementation, process technology and workload. This energy value includes peripheral logic and memory access, as well as $\sim$25~fJ for the MAC operation alone (from Ref.~\cite{horowitz20141}, scaled to a 7~nm node~\cite{stillmaker2017scaling}). The energy of our single-shot ONN, on the other hand, is on the order of $\sim$10~fJ/MAC, including a source wall-plug efficiency of 10\%~\cite{kuramoto2018high,skalli2022photonic}, PD responsivity of 0.2~A/W~\cite{fahs2017design}, TIA sensitivity of 1~$\upmu$A at 1~GHz~\cite{mehta201612gb}, LCoS SLM power consumption of <10~W and TIA~\cite{mehta201612gb}, ADC~\cite{jonsson2011empirical,tripathi20138}, DAC~\cite{morales2022analysis,tripathi20138} and nonlinearity~\cite{Sze2017-qg,jouppi_tpu} energies of 1~pJ per operation. This overall energy per MAC, including electrical-to-optical and optical-to-electrical conversion, is therefore similar to the cost of one digital electronic MAC, before even considering the expensive electronic data movement in digital accelerators.

The chip area of our proposed near-term system ($\sim$30~mm$^2$), dictated primarily by the size of the weighting pixels, is similar to the area consumed by 1,000~$\times$~1,000 electronic MAC units without peripheral logic. In the single-shot ONN, the chip area is primarily dictated by the size of the weighting elements in the receiver array. In a digital electronic weight-stationary array, each MAC unit has an area of (5-8~$\upmu$m)$^2$ (8-bit multiplier~\cite{saadat2018minimally,shoba2017energy,johnson2018rethinking} scaled to a 7~nm node~\cite{stillmaker2017scaling}), for a total million-element area of 25-64~mm$^2$.

For more details on the latency, throughput, energy and area estimates, see Supplementary Note~2.

\section*{Backmatter}

\begin{backmatter}

\bmsection{Acknowledgments}
We thank Prof. David A. B. Miller and Hugo Larocque for helpful discussions, Ian Christen for support with spectral measurements and the MITRE Corporation for the loan of one LCoS SLM.
This work is supported by NTT Research Inc., the National Science Foundation EAGER program (CNS-1946976) and the U.S. Army Research Office through the Institute for Soldier Nanotechnologies (ISN) at MIT (W911NF-18-2-0048). L.B. is supported by the Natural Sciences and Engineering Research Council of Canada (PGSD3-517053-2018), A.S. is supported by the National Science Foundation GRFP (1745302), C.P. is supported by the Hertz Foundation Elizabeth and Stephen Fantone Family Fellowship, and S.T.-M. is supported by the VATAT postdoctoral scholarship for quantum technology.

\bmsection{Data availability}
Data underlying the results presented in this paper are available upon reasonable request.

\bmsection{Code availability}
Relevant code is available upon reasonable request.

\bmsection{Competing interests}
D.E. is an advisor to LightMatter. Patent application No. 17/673,268 filed on 2022-02-16 by Liane Bernstein, Alexander Sludds and Dirk Englund.

\bmsection{Author contributions}
L.B., A.S. and D.E. developed the concept. C.P. and S.T.-M. wrote the holography software used to generate the fan-out pattern. R.H. and D.E. supervised the research and guided the experiments. L.B. built the experimental system with coding help from A.S. L.B. conducted the experiments, wrote the simulation software and wrote the manuscript with input from all the authors.

\bmsection{Supplementary information}
See Supplementary Information for supporting content.

\end{backmatter}


\bibliography{sample}

\end{document}


\maketitle

\section{Experiment: detailed description}

\begin{figure}[htpb]
    \centering
    \includegraphics[width=.6\linewidth]{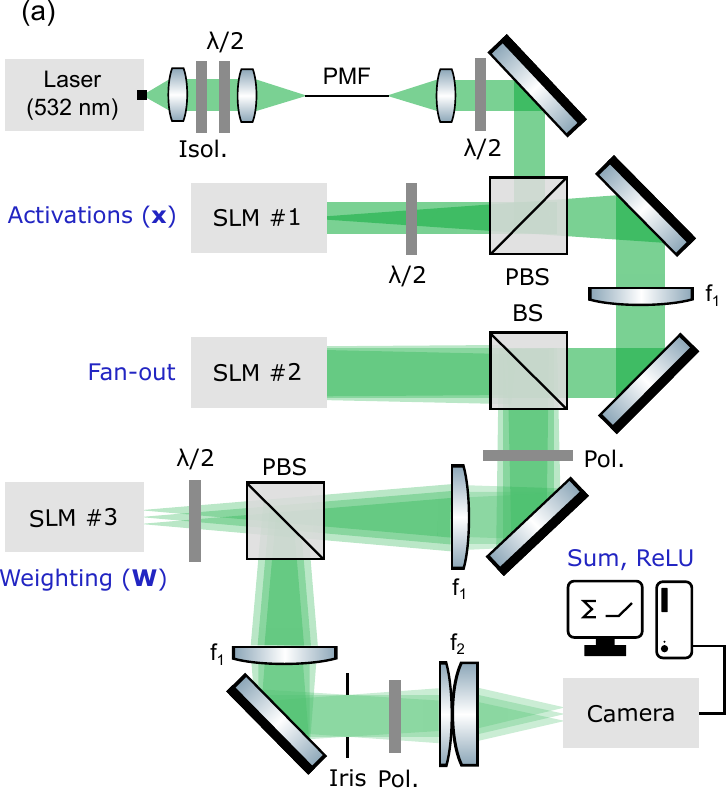}
    \caption{Complete system elaborating on Fig~3(a) from main text, including isolator (Isol.), single-mode polarization-maintaining fiber (PMF), iris and polarizers (Pol.) to reduce stray light. Focusing lens f$_2$ is in fact two achromatic lenses, with individual focal lengths of 750~mm and 180~mm.}
    \label{fig:detailed_experiment}
\end{figure}

This section provides more details on our experimental implementation, including added components to reduce stray light (iris, polarizers) --- see Fig.~\ref{fig:detailed_experiment}. The first spatial light modulator (SLM, Meadowlark, AVR Optics P1920-400-800-HDMI-T, pixel width 9.2~$\upmu$m) displays an input image \textbf{x} by modulating the phase of the polarization component of the incident collimated beam that is parallel to the extraordinary axis. (The polarization component along the ordinary axis is unchanged.) Because the light incident on this SLM is polarized at 45$\degree$ after rotation by the half-wave plates, the SLM effectively performs pixel-wise polarization rotation of the beam. A polarizing beamsplitter (PBS) then rejects the unrotated polarization from the SLM. The input activations are thus encoded into optical intensities through this `amplitude-mode' operation of a `phase-only' SLM. To display the input activations in the first layer of the DNN, we use every second pixel of the first SLM and turn every other pixel `off' such that the input pattern is effectively multiplied by a grating. We then block the zero order in the Fourier plane to reduce background, such as reflection from the backplane. In subsequent layers with fewer input activations (shorter \textbf{x}), we use every eighth pixel of the SLM to further reduce crosstalk. The incident light on the second SLM (Hamamatsu X10468-04, pixel width 20~$\upmu$m) is horizontally polarized (along the extraordinary axis); this SLM is in the Fourier plane and adds a variable phase delay to each pixel to impart a fan-out phase pattern. We used the weighted Gerchberg-Saxton algorithm (without camera feedback) to determine the phase pattern~\cite{Kim2019-gs}, which only needed to be calculated once, as it is independent of the weight and input activation values. The third SLM (Meadowlark, AVR Optics P1920-400-800-HDMI-T, pixel width 9.2~$\upmu$m), in the image plane, is used in `amplitude mode', similarly to the first SLM. Telescopes of achromatic lenses (Thorlabs ACT508-250-A, AC508-180-A, and ACT508-750-A) transmit the replicated input activations to the image planes for 1:1 mapping from SLM~\#1 to SLM~\#3 to the camera (Thorlabs DCC3240M, pixel width 5.3~$\upmu$m). The lens positions along the optical axis are fine-tuned with linear stages. A MATLAB program performs all the control and processing on a digital electronic computer. The supercontinuum source in the throughput experiments is the SuperK EXW-12 from NKT with a VARIA tunable filter.

\section{Latency, throughput, energy and chip area of a near-term system}

\begin{figure}[htpb]
    \centering
    \includegraphics[width=.8\linewidth]{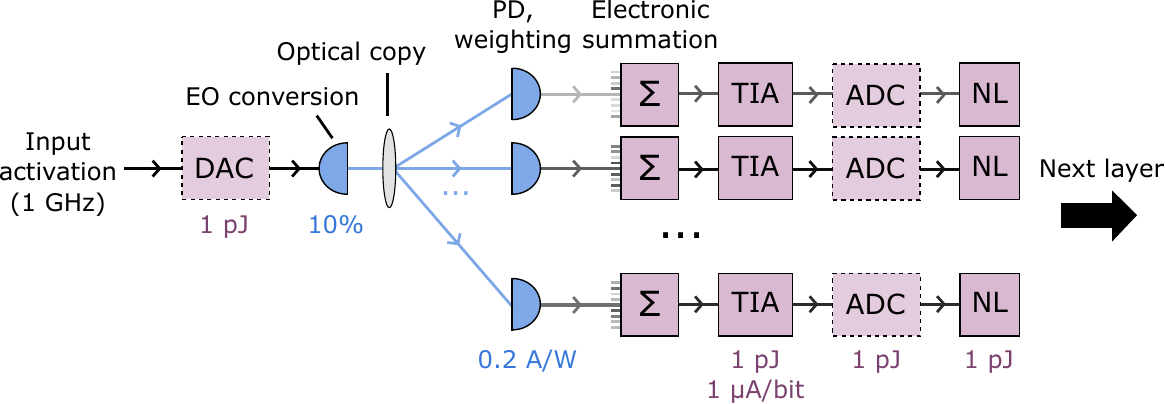}
    \caption{Path of one input activation through the single-shot ONN in an optimized setup. A VCSEL or $\upmu$LED converts the signal from the electronic to optical domain (EO conversion) and the optical copy is a reconfigurable diffractive optical element. Each photodetector (PD) includes a weighting element and is electrically connected to $K$ other PDs for analog electronic summation. A transimpedance amplifier (TIA) reads out the analog output signal from each block, which the DAC then converts to the digital domain. The nonlinearity is a simple comparator. Each of the $K$ input activations goes through these processing steps simultaneously such that after one pass, the matrix-vector product is complete.}
    \label{fig:data_path}
\end{figure}

In this section, we describe the latency, throughput, energy and chip area of a near-term, CMOS-compatible implementation of our single-shot ONN. This proposed system, as shown in Fig.~\ref{fig:data_path} and described in the main text, includes digital-to-analog converters (DAC) and a high-speed source array with $K$ elements, LCoS SLMs for fan-out and weighting, photodetectors (PDs) for optical-to-electrical (OE) conversion, passive, analog electronic summation, and for each block, amplification by a transimpedance amplifier (TIA), analog-to-digital conversion (ADC) and a nonlinearity (NL). We assume a matrix size of 1 million elements ($N=K=1$,000), as megapixel cameras and SLMs are readily available. The latency of the system is defined by the sum of the latencies of each of the components encountered by one input activation, where the DAC, light source, TIA, ADC and NL take $\sim$1~ns each (a standard computer clock operates at $\sim$GHz), and the photon time of flight is $<$10~ns. Thus, the system operates with a latency on the order of $\sim$10~ns for a full matrix-vector computation.

As mentioned in the main text, computation is pipelined such that the throughput is defined by the slowest component, yielding $10^{6}$ MACs per $\sim$1~ns, i.e., petaMAC/s for $N=K=1,000$.

The energy consumption for a complete matrix-vector multiplication that comprises $N\times K$ MACs is the sum of the DAC, SLM, TIA, ADC and nonlinearity energies, plus the photon energy required to discriminate 256 levels on the TIA:
\begin{align}
    E_\text{total}\text = N\cdot \tfrac{1}{\eta_\text{Laser} \cdot \eta_\text{PD}} \cdot 2^{n_b} \cdot \xi \cdot t + K\cdot E_\text{DAC} + 2\cdot E_\text{SLM}~+ \nonumber \\ 
    N\cdot (E_\text{TIA} + E_\text{ADC} + E_\text{NL})
\end{align}
where $\eta_\text{Laser}\approx$~10\% is the source wall-plug efficiency~\cite{kuramoto2018high,skalli2022photonic}, $\eta_\text{PD}\approx$~0.2~A/W is the PD responsivity~\cite{fahs2017design}, $n_b$~=~8~bits, $\xi \approx$~1~$\upmu$A is the TIA sensitivity at 1~GHz~\cite{mehta201612gb}, $t$~=~1~ns is the integration time, i.e., computer clock cycle time, $E_\text{SLM}$ is the energy consumed by the SLM in one clock ($<$10~nJ) and each of the remaining component energies (TIA~\cite{mehta201612gb}, ADC~\cite{jonsson2011empirical,tripathi20138}, DAC~\cite{morales2022analysis,tripathi20138}, NL~\cite{Sze2017-qg,jouppi_tpu}) are 1~pJ per operation. For $N = K = 1,000$, the energy per MAC ($E_\text{total}/(N\cdot K)$) is therefore on the order of $\sim$10~fJ/MAC. These parameters and calculations are summarized in Table~\ref{tab:parameters}.

\begin{table}
\centering
\footnotesize
\begin{threeparttable}
\caption{Parameters in energy calculation}
\begin{tabular}{ccccc} 
\hline
Symbol & Parameter & Value\tnote{2} & Fan-out & Energy/MAC\tnote{4} \\
\hline
$\eta_\text{Laser}$ & Laser wall-plug efficiency & 10\%~\cite{skalli2022photonic,kuramoto2018high} & & \\
$\eta_\text{PD}$ & PD responsivity & 0.2~A/W~\cite{fahs2017design} & & \\
$n_b$ & Effective number of bits & 8 & & \\
$\xi$ & TIA sensitivity & 1~$\upmu$A~\cite{mehta201612gb} & & \\
$t$ & Clock cycle time & 1~ns & & \\
$E_\text{DAC}$ & Energy of one DAC conversion & 1~pJ~\cite{morales2022analysis},\cite{tripathi20138}\tnote{3} & $N$ & 1~pJ/$N\rightarrow$ 1~fJ/MAC \\
$E_\text{SLM}$ & Energy consumed by LCoS SLM & ($<$10~W)$\cdot t$ & $N\times K$ & $<$10~nJ/$(N\times K)\rightarrow$ $<$10~fJ/MAC \\
$E_\text{optical}$ & Optical energy in one block\tnote{1} & $\tfrac{1}{\eta_\text{Laser} \cdot \eta_\text{PD}} \cdot 2^{n_b} \cdot \xi \cdot t \approx 10$~pJ & $K$ & 10~pJ/$K\rightarrow$ 10~fJ/MAC \\
$E_\text{TIA}$ & Energy of TIA & 1~pJ~\cite{mehta201612gb} & $K$ & 1~pJ/$K\rightarrow$ 1~fJ/MAC \\
$E_\text{ADC}$ & Energy of one ADC conversion & 2~pJ~\cite{jonsson2011empirical,tripathi20138} & $K$ & 2~pJ/$K\rightarrow$ 2~fJ/MAC \\
$E_\text{NL}$ & Energy of nonlinearity & $<$1~pJ~\cite{Sze2017-qg,jouppi_tpu} & $K$ & $<$1~pJ/$K\rightarrow$ $<$1~fJ/MAC \\
\hline
$E_\text{total}$ & Energy of full system & & & $\sim$10~fJ/MAC \\
\hline
\end{tabular}
\begin{tablenotes}
\item [1]Optical energy for $2^{n_b}$ distinguishable levels by the TIA.
\item [2]Demonstrated in the literature.
\item [3]DAC within ADC.
\item [4]Assuming $N=K=1$,$000$.
\end{tablenotes}
\label{tab:parameters}
\end{threeparttable}
\end{table}

The total chip area of the system is calculated in Table~\ref{tab:area}. The component areas have been demonstrated experimentally in the literature, in CMOS technology nodes larger than the state of the art -- therefore, the TIA, ADC, NL, DAC and VCSEL areas could likely be further miniaturized. The weighting elements, on the other hand, will also be limited by optical spot size. The diffraction limit is $<\upmu$m for green light, but a pixel size of a few $\upmu$m is more reasonable to maintain low crosstalk in a real system that includes imperfections (i.e., aberrations and misalignment).

\begin{table}
\centering
\footnotesize
\begin{threeparttable}
\caption{Chip area}
\begin{tabular}{cccc} 
\hline
Element & Area (mm$^2$) & Number of elements & Total area (mm$^2$) \\
\hline
Weighting element\tnote{1} & $1.4\cdot 10^{-5}$~\cite{holoeye} & $10^6$ & 14 \\
TIA & $0.0022$~\cite{mehta201612gb} & $10^3$ & 2.2 \\
ADC & 0.0016~\cite{kull201728} & $10^3$ & 1.6 \\
NL & 0.001~\cite{vijaya2016low} & $10^3$ & 1 \\
DAC & $<$0.0016~\cite{kull201728}\tnote{2} & $10^3$ & $<$1.6 \\
VCSEL & .01~\cite{kuramoto2019watt}\tnote{3} & $10^3$ & 10 \\
\hline
Area of full system & & & 30 \\
\hline
\end{tabular}
\begin{tablenotes}
\item [1]Includes PD (e.g., PDs in Refs.~\cite{Miller2017-qd,latif2009low}), since weighting element is placed on top of each PD.
\item [2]DAC within ADC.
\item [3]Aperture of 6~$\upmu$m, pitch of 100~$\upmu$m in demonstrated array.
\end{tablenotes}
\label{tab:area}
\end{threeparttable}
\end{table}

\section{Negative weights}
A number of solutions can implement negative weighting. For example, a second wire can connect all the PDs in a block; charge from the negatively weighted pixels would be directed into this second wire with an analog switch. The output from the `negative' wire can then be subtracted from the output of the `positive' wire. Because this subtraction only occurs once per block, its cost is amortized by a factor of $K$ and is therefore small with respect to the other costs of the system. Another possibility for negative weighting is to use two PDs per receiver pixel, where one PD pushes charge into the block's wire in the case of a positive weight, and another pulls charge in the case of a negative weight. The weight value of the unused photodetector is set to maximum extinction. Lastly, the weights can be shifted to all positive values, as described by Wang et al.~\cite{wang2022-1photon}.

\section{Calibration and image processing}

The LCoS SLM calibrations from the manufacturers are set to map the input values of 0 to 255 linearly to a 0 to 2$\uppi$ phase shift for normally incident 532~nm light. Since we use the SLMs from Meadowlark/AVR Optics (i.e., SLMs \#1 and \#3) to modulate amplitude and not phase, the output light intensity then varies sinusoidally with input value and requires recalibration. This SLM model has 2048 voltage settings in hardware with $>$2$\uppi$ phase modulation. In the calibration step, we display a uniform array at one voltage value, then average the outputs to obtain one output value per input voltage. We then step through the input voltage values linearly through time, and fit the inputs versus averaged outputs with a 9th-order polynomial. We use the resulting lookup table as a global calibration to map desired output intensities to SLM input values. The latter are then restricted to 7-8 bits of precision depending on the intensity region since there are multiple maximum to minimum output intensity cycles within the 2048 available input voltage values. Furthermore, because a wide region of SLM~\#3 is illuminated, local nonuniformities will cause different blocks (subimages) to require slightly different calibrations to achieve a linear weight display. Therefore, we calculate a refined fit per subimage, where we once again display a uniform array at each frame, but here, step through the display values from the global lookup table. We then fit 8th-order polynomials to the inputs versus averaged outputs per subimage and adjust the displayed weight values accordingly.

We also perform simple processing of the output images from the camera. To reduce the impact of stray light in the system, we acquire a background with SLMs~\#1 and \#3 set to all zeros, and subtract this background from every output frame. We also perform $2\times2$ pixel binning. Furthermore, the fan-out pattern displayed on SLM~\#2 does not yield subimages of equal average intensities on the camera. Therefore, we also acquire a smoothed and background-subtracted calibration map by setting all activations and weights to a constant value, used for normalization in our experiments. In layer 1, the outputs are also divided by the mean intensity per subimage of 100 randomly selected images from the MNIST validation set. The normalization process reduces the effective number of bits for some subimages since we do not make use of the full dynamic range of the camera. The spot uniformity could be improved by adjusting the pattern on SLM~\#2 with a feedback algorithm~\cite{Kim2019-gs}, eliminating the need for normalization. In an optimized implementation, the background subtraction can be implemented with a bias voltage.

\section{Measurement of filtered supercontinuum laser spectra}

We corrected the measured supercontinuum spectra for the wavelength response of our experimental setup, which has low transmission for wavelengths $<$460~nm or $>$560~nm, primarily due to the dielectric mirror of SLM~\#2. To determine our system's response function, we first set the VARIA filter attachment on the SuperK to the narrowest bandwidth. We then swept its center wavelength (every 10~nm from 450~nm to 630~nm) and measured the summed intensity on the camera with SLMs~\#1 and \#3 set to a uniform display. We performed a similar measurement when we connected the supercontinuum laser to the spectrometer, where we measured and integrated the filtered spectrum at the narrowest bandwidth setting for different center wavelengths. We then estimated our system's wavelength response as the summed camera intensity per wavelength divided by the spectrometer's summed intensity per wavelength. We could then calculate the corrected spectra by multiplying the broad-bandwidth spectra (measured on the spectrometer) by this system response (interpolated).


\bibliography{supplement}